\documentclass[journal=apchd5,manuscript=article]{achemso}

\usepackage[version=3]{mhchem} 
\usepackage[utf8]{inputenc}
\usepackage{textcomp}  

\usepackage{enumitem}
\usepackage[usenames,dvipsnames]{color}

\newcounter{figtodo_c}

\renewcommand\Im{\operatorname{Im}}

\author{Taavi Rep\"an}
\affiliation{Institute of Physics, University of Tartu, Ravila 14c, Tartu 50411, Estonia}
\alsoaffiliation{DTU Fotonik, Technical University of Denmark, {\O}rsteds pl. 343, 2800 Kongens Lyngby, Denmark}

\author{Andrei V. Lavrinenko}
\affiliation{DTU Fotonik, Technical University of Denmark, {\O}rsteds pl. 343, 2800 Kongens Lyngby, Denmark}

\author{Sergei V. Zhukovsky}
\affiliation{DTU Fotonik, Technical University of Denmark, {\O}rsteds pl. 343, 2800 Kongens Lyngby, Denmark}
\email{sezh@fotonik.dtu.dk}

\title[Dark-field hyperlens]
  {Dark-field hyperlens: Super-resolution imaging of weakly scattering objects}

\abbreviations{IR,NMR,UV} 

\begin{document}

\begin{tocentry}

  We propose and numerically demonstrate a technique for subwavelength imaging based on a metal-dielectric multilayer hyperlens designed in such a way that only the large-wavevector waves are transmitted while all propagating waves from the image area are blocked by the hyperlens. As a result, the image plane only contains scattered light from subwavelength features of the objects and is free from  background illumination. Similar in spirit to conventional dark-field microscopy, the proposed {\em dark-field hyperlens} is promising for optical imaging of weakly scattering subwavelength objects, such as optical nanoscopy of label-free biological objects.

\end{tocentry}

\begin{abstract}
  We propose and numerically demonstrate a technique for subwavelength optical imaging based on a metal-dielectric multilayer hyperlens designed in such a way that only the large-wavevector waves are transmitted while all propagating waves from the image area are blocked by the hyperlens. As a result, the image plane only contains scattered light from subwavelength features of the objects and is free from  background illumination. Similar in spirit to conventional dark-field microscopy, the proposed {dark-field hyperlens} is numerically shown to enhance the subwavelength image contrast by more than two orders of magnitude. These findings are promising for optical imaging of weakly scattering subwavelength objects, such as optical nanoscopy of label-free biological objects.
\end{abstract}

\section{Introduction}

The recent decade in modern materials science has featured the advent of optical metamaterials, where the role of known, ordinary constituents of matter (atoms, ions, or molecules) is bestowed upon artificial ``meta-atoms''---nanosized objects purposely designed to have the desired optical properties \cite{bookShalaev}.  If the meta-atoms are much smaller than the wavelength of light interacting with them, then the meta-atom assembly, or an artificial composite metamaterial, would exhibit the desired properties macroscopically. The elegance of the metamaterials concept lies in the nearly limitless potential variety of meta-atom shapes and compositions, which surpasses the variety of naturally occurring atoms, molecules and crystals (and, in turn, of natural materials).

\begin{figure}
\includegraphics[width=1\columnwidth]{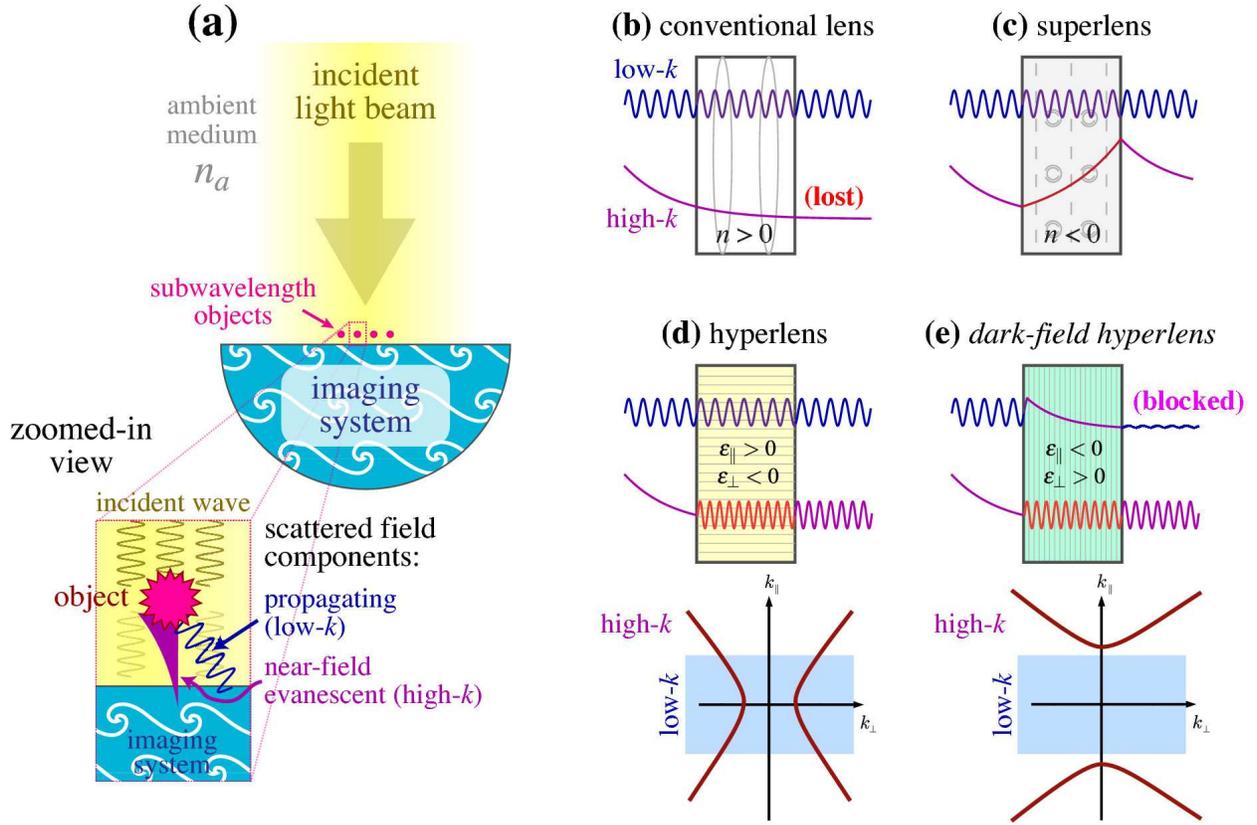}
\caption{(a) Overview of an optical imaging system with subwavelength resolution, schematically showing propagating low-$k$ ($k<n_a \omega/c$) and evanescent high-$k$ ($k>n_a \omega/c$) components of radiation scattered off subwavelength objects. 
(b--e) Schematics of how 
information contained in these components   
passes through different types of imaging systems. (b) {\em Conventional optical lens}: only the propagating low-$k$ components are transmitted while the high-$k$ evanescent waves carrying near-field information are lost, resulting in a blurred, diffraction-limited image. (c) {\em Superlens}\cite{pendry2000negative}: high-$k$ evanescent waves are amplified and the near-field information is recovered, enabling a subwavelength image, which nevertheless remains in the near-field and cannot be reproduced by conventional optics. (d) {\em Hyperlens}\cite{jacob2006optical}: high-$k$ evanescent waves are converted to propagating waves using an HMM\cite{poddubny2013hyperbolic}; the resulting subwavelength image can therefore be seen in the far field, however its contrast against the background illumination will be poor if the object is weakly scattering. (e) {\em Proposed dark-field hyperlens}: using a modified kind of HMM \cite{ourReviewPNFA} similarly couples the high-$k$ evanescent waves to the far field but blocks the low-$k$ propagating waves, filtering out the background illumination and allowing the subwavelength image contract to be drastically enhanced. For the two kinds of hyperlenses, the example dispersion relations are shown as insets in (d) and (e).  }
\label{FIG:Lenses}
\end{figure}

The hallmark success of optical metamaterials is the design of artificial materials with optical properties that do not exist in naturally occurring media, such as negative refractive index \cite{shalaev2007optical}. Such negative-index media   came out as seminal to the metamaterials field because of the vision of the ``perfect lens'' \cite{pendry2000negative},  where a slab of an artificial material with $n = -1$ (a ``super-lens'') would focus light tighter than diffraction would allow in a conventional optical system. Even though this ``perfect lens'' dream, which hinges on the existence of lossless and isotropic negative-index metamaterials, may never come true, it did give birth to the entire field of study with several successful experimental demonstrations of subwavelength imaging \cite{SWexp05,liu2007experimental,aydin2007subwavelength,xiong2007two}. It was understood that the operating principle of the superlens is its ability to transmit, rather than to lose, the near-field information about a subwavelength object \cite{zhang2008superlenses}, as seen in \ref{FIG:Lenses}(a--c). 

Later studies have shown that so-called {\em hyperbolic metamaterials} (HMMs)\cite{poddubny2013hyperbolic}, which are extremely anisotropic media that are metal-like along some coordinate axes and dielectric-like along others, make it possible to do more---to convert the near field of an object into a set of propagating waves to be later imaged by conventional means. This concept of the hyperlens \cite{jacob2006optical} \ref{FIG:Lenses}(d),  followed by experimental demonstrations\cite{liu2007far}, showed that subwavelength imaging could be far closer to reality than one would assume after the initial disappointment in the superlens. Subsequent theoretical studies have further shown that a broad variety of structures, including metal-dielectric  multilayers\cite{ourReviewPNFA},  possess the necessary requirements to function in a hyperlens. Such multilayers are of much simpler geometry than metamaterials commonly required to achieve negative refractive index. 

The existing designs of the hyperlens recover information from both propagating (low-$k$) waves with $k<n_ak_0$ and evanescent (high-$k$) waves with $k>n_ak_0$ (where $n_a$ is the ambient refractive index and $k_0=\omega/c$). Such an approach is undoubtedly the best way of maximizing the output from the object to create the brightest possible subwavelength image. However, this approach has a serious downside: any propagating waves that exist in the object area but do not originate form the object (such as incident or stray light) would be transmitted, creating strong background in the image area. It is for this reason that existing demonstrations of the hyperlens focus on examples where background radiation can be eliminated. This is done either by imaging a subwavelength pattern in a metal screen \cite{liu2007far,rho2010spherical,cheng2013breaking} that covers the entire lens and blocks all incident light, or else by using self-illuminating objects such as fluorescent centers \cite{li2009experimental,andryieuski2012graphene}. In a scenario when objects to be imaged are weakly scattering and have to be illuminated by external light, as is very relevant in label-free biological imaging, a conventional hyperlens would be nearly useless because the resulting image would have extremely low contrast. 

In this paper, we propose an alternative hyperlensing concept which is free from this downside and can provide high-contrast subwavelength images of weakly scattering objects. The proposed device only transmits high-$k$ waves while blocking all propagating radiation in the sample area (\ref{FIG:Lenses}d), be it from the object itself or from elsewhere. The resulting image would therefore only contain information coming from subwavelength features of the sample, providing a much greater contrast than the conventional hyperlens. The proposed device relates to the conventional hyperlens in the same way as dark-field microscopy relates to conventional optical microscopy. Therefore we have termed the proposed device the {\em dark-field hyperlens} (DFHL), and the proposed concept, {\em dark-field superresolving optical microscopy}. Unlike an earlier work of 2012 by H. Benisty, where the term dark-field hyperlens was first coined\cite{benisty2012dark}, the filtering of the background radiation is done by the imaging hyperlens itself rather than by using a second hyperlens to excite the sample in the confocal geometry. As a result, the proposed device geometry is much simpler and within easier reach of modern fabrication facilities.

It should be stressed that beating the diffraction limit in the optical microscopy would constitute a major breakthrough in biological imaging because it neither requires special sample fixation and preparation techniques (such as electron-microscopy and scanning-microscopy methods) nor relies on special labeling methods (such as STED-like approaches\cite{STEDreview}). As a result, the proposed dark-field superresolving optical microscopy can be used to obtain dynamic real-time images of weak-contrast subwavelength objects. Being able to see and investigate dynamic processes involving very small biological agents and macromolecules would be truly enabling to modern life sciences, as confirmed by the ongoing scientific efforts in search for such a technique\cite{Ayas:2013rt,wong2013optical,yang2014super}. Hence, the impact of such subwavelength optical microscopy, or nanoscopy, that would be high-resolution, high-contrast, fast, and non-destructive to the specimen---such as the dark-field hyperlens can potentially provide---can potentially be on par with the impact of the original invention of an optical microscope on biology several centuries ago. The accompanying possibility to reverse the operation of a hyperlens in order to selectively excite an object on a subwavelength scale may bring about an even greater advance in experimental biology, making it possible not only to observe the functioning of nanoscale biological agents, but also to actively interfere with them.


The paper is organized as follows. Section 2 introduces the principle of the DFHL, providing analytical guidelines to its operation. Section 3 deals with the numerical demonstration using planar-geometry HMM as a toy-model or ``poor man's'' hyperlens, which provides subwavelength imaging without magnification. Section 4 moves on to provide a numerical demonstration of a cylindrical-geometry DFHL, confirming subwavelength imaging functionality with magnification rate of 2.5 for objects down to 300 nm apart. Finally, Section 5 summarizes the paper.


\section{Operating principle of the dark-field hyperlens}

The primary physical concept behind subwavelength imaging in plasmonic metamaterials---the operating principle of the hyperlens---is the idea that a medium with extreme anisotropy, such that the components of the permittivity tensor have different signs, supports propagating waves with very large wave vectors. Indeed, recalling the dispersion relation for the extraordinary (TM-polarized) wave in a uniaxial birefringent medium, we get
\begin{equation}
k_0^2=\frac{\omega^2}{c^2}=\frac{k_\parallel^2}{\epsilon_\perp}+\frac{k_\perp^2}{\epsilon_\parallel},
\label{eq:disp}
\end{equation}
where $\epsilon_\perp$ and $\epsilon_\parallel$ are components of the dielectric permittivity tensor $\hat\varepsilon = \mathrm{diag}(\epsilon_\parallel,\epsilon_\parallel,\epsilon_\perp)$; $k_\perp$ and $k_\parallel$ are respective components of the wave vector. We see that for normal isotropic or weakly birefringent media the solutions of \eqref{eq:disp} in the $k$-space represent  bounded shapes (ellipsoids), providing an upper limit on the possible values of $k$ for propagating waves. In contrast, if the optical anisotropy is so strong that $\epsilon_\perp$ and $\epsilon_\parallel$ are of different signs, then the solutions of \eqref{eq:disp} change topology from bounded ellipsoids to unbounded hyperboloids, thus supporting propagating solutions with theoretically infinite wave vectors\cite{smith2004partial}. 

This fact had remained largely a theoretical curiosity until such hyperbolic dispersion could actually be realized for optical waves in HMMs---subwavelength metal-dielectric structures with rather simple geometries such as nanorod arrays and multilayers \cite{poddubny2013hyperbolic,ourReviewPNFA}. It was shown that even though ``infinitely large wave vectors'' proved to be an idealization\cite{ourHMMPRA}, HMMs  can indeed support propagating plasmonic waves with very large $k$-vectors\cite{ourPoleExpOE}.

As a result, an HMM can transform high-$k$ waves with $k_\parallel > n_a k_0$, which are evanescent in the ambient medium with refractive index $n_a$, to high-$k$ waves that can propagate through the metamaterial. To see how this gives rise to subwavelength imaging properties, we recall that any object's scattered field can be decomposed into a series of plane waves spanning the entire range of $k$. The relation between the spatial representation of the object $f(x,y)$ and its image $g(x,y)$ in the Fourier optics approach is 
\begin{equation}
g(x,y)=\frac{1}{(2 \pi)^2}
\iint_{-\infty}^{\infty} dk_x dk_y H(k_x,k_y)e^{-i(k_x x+k_y y)}
\iint_{-\infty}^{\infty} dx' dy' f(x,y) e^{i(k_x x'+k_y y')} .
\label{eq:fourier2D}
\end{equation}
Here $H(k_x,k_y)$ is the transfer function of the imaging system, and it can be seen from the properties of Fourier transformation that $g(x,y)=f(x,y)$ (the image is perfect) if $H(k_x,k_y)=1$. In conventional optics, propagation of light over some distance $d$ in the ambient medium introduces a low-pass filter in the $k$-space, resulting in the transfer function $H_0(k_x,k_y)\propto \exp(-idk_\perp) = \exp\left({-id\sqrt{n_a^2 k_0^2-k_\parallel^2}}\right)$ with $k_\parallel^2=k_x^2+k_y^2$. Therefore, if the object's Fourier image is significantly extended into the area with $k_\parallel > n_a k_0$, or in other words, if the size of the object is smaller than the wavelength of light $\lambda=2\pi n_a/k_0$, then all the components with $k_\parallel > n_a k_0$ (the near-field information) are lost (see \ref{FIG:Lenses}(b)), and the image becomes blurred.

This low-pass filtering is overcome in HMMs, where \eqref{eq:disp} results in the expression 
\begin{equation}
k_\perp = \sqrt{\epsilon_\parallel k_0^2-(\epsilon_\parallel/\epsilon_\perp)k_\parallel^2},
\label{eq:kperp}\end{equation}
which can remain real for very large $k_\parallel$ because $\epsilon_\parallel/\epsilon_\perp<0$. This makes the transfer function such that the loss of near-field information is prevented  (see \ref{FIG:Lenses}(d)). Thus, placing an HMM close to the object facilitates superresolution imaging. 

This idea, combined with the use of curvilinear geometry so that high-$k$ waves inside the metamaterial can be further coupled to outside propagating waves and imaged by a conventional lens, was put to use in the practical realization of the hyperlens\cite{liu2007far}. Later studies followed with experimental demonstration of a spherical hyperlens at visible frequencies \cite{rho2010spherical} and the design of an all-dielectric hyperlens\cite{jiang2013broadband}, as well as with applications to other platforms such as terahertz  \cite{andryieuski2012graphene} and acoustic  \cite{li2009experimental} waves. 

Depending on the signs of the eigenvalues in the dielectric permittivity tensor, hyperbolic media can be classified as either type I ($\epsilon_\perp < 0 < \epsilon_\parallel$) or type II ($\epsilon_\parallel < 0 < \epsilon_\perp$)\cite{Guo2012}. In the periodic metal-dielectric multilayer geometry, the effective permittivity components can be obtained from the Maxwell-Garnett homogenization approach with \cite{ourHMMPRA}
\begin{equation}
\epsilon_\parallel = \rho \epsilon_m + (1-\rho) \epsilon_d, \quad
\epsilon_\perp = \left[\rho \epsilon_m^{-1} + (1-\rho) \epsilon_d^{-1} \right]^{-1},
\label{eq:epscomponents}
\end{equation}
where $\rho=d_m/(d_m+d_d)$ is the filling fraction of the metal. Therefore, by choosing the thicknesses of metal and dielectric layers in the stack ($d_m$ and $d_d$), as well as the permittivities of metal and dielectric ($\epsilon_m$ and $\epsilon_d$), one can design a structure that is effectively a type-I or type-II HMM.

\begin{figure}
\includegraphics[width=1.00\columnwidth]{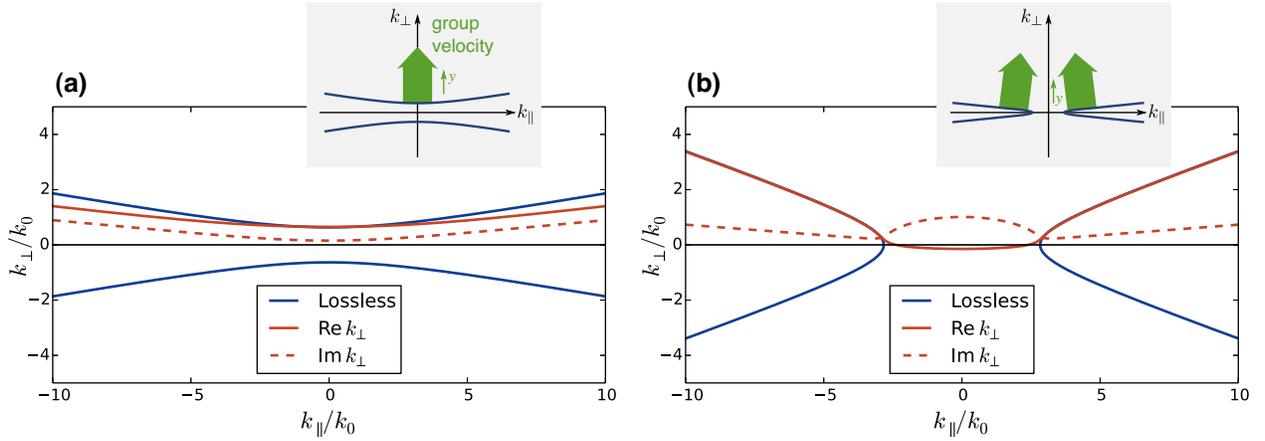}
\caption{Two-dimensional dispersion relation for (a) type I hyperbolic metamaterial (dielectric permittivity tensor compoments being $\epsilon_\parallel = 0.36$, $\epsilon_\perp = -13.31$) and (b) type II hyperbolic metamaterial ($\epsilon_\parallel = -1.06$, $\epsilon_\perp = 8.09$) The insets schematically show the direction of the group velocity for waves in certain parts of the $k$-space; for the type I hyperbolic metamaterial, it is possible that the majority of lower-$k$ waves propagate in the $y$-direction (the canalization regime).
}
\label{FIG:Dispersion}
\end{figure}

The key difference between the two types of HMMs can be seen from \eqref{eq:disp}: the dispersion contour in the wave vector space has a different topology, forming either one connected or two unconnected hyperboloidal surfaces (see \ref{FIG:Dispersion} for example structures). As a result [see \eqref{eq:kperp}], type-I HMMs support bulk propagating waves with {\em any} value of the tangential wave vector $k_\parallel$, and act like dielectrics for TE-polarized (ordinary) waves. In contrast, type-II HMMs only support TM-polarized bulk propagating waves for $k_\parallel>k_\text{cutoff}=\sqrt{\epsilon_\perp}k_0$, exhibiting effective metallic properties for lower $k_\parallel$ as well as for the TE-polarized waves.

Consider now a scenario where there is a subwavelength object placed in front of an HMM-based hyperlens and illuminated by external light. At first sight, a type-I HMM is much better suited for the design of the hyperlens because it transmits both propagating components of the scattered radiation (with $k_\parallel < n_a k_0$) and all the evanescent components (with $k_\parallel > n_a k_0$, where $n_a$ is the refractive index of the ambient medium). As a result, subwavelength image with maximum resolution and good brightness can be formed. However, a type-I HMM would transmit the propagating low-$k$ components originating not only from the object, but also from other sources, such as incident or stray light. Therefore, if the object to be imaged is weakly scattering (as is the case with most subwavelength objects made of dielectrics, such as all biological objects), the resulting image would have very low contrast, rendering such a hyperlens extremely difficult in use. 

What we propose as the main idea of this paper is to modify the hyperlens in such a way that a type-I HMM is replaced with a type-II one designed so that $k_\text{cutoff}$ exceeds $n_a k_0$. This would filter all low-$k$ propagating components, so that only the high-$k$ waves stemming from the objects to be imaged would make it to the image area. Such a device would in some sense represent the ``inverse'' of the conventional optics [compare \ref{FIG:Lenses}(b) and (e)], introducing high-pass rather than low-pass filtering in the transfer function of the hyperlens. 

\begin{figure}[tbh]
\includegraphics[width=1.0\columnwidth]{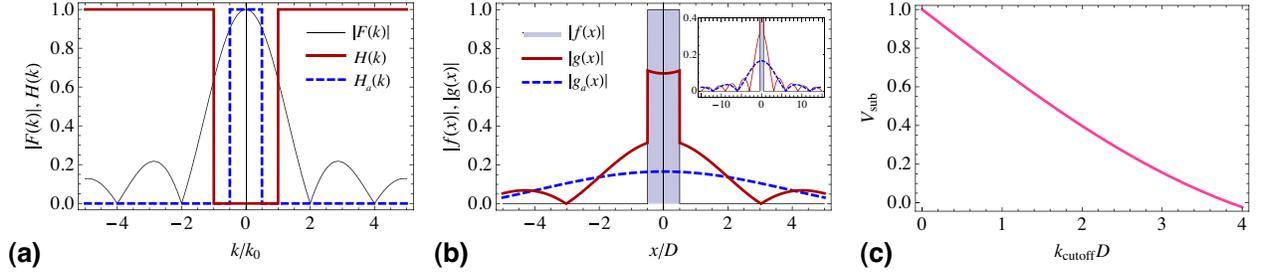}
\caption{(a) The Fourier transform $F(k)$ of a subwavelength object given by $f(x)=\Pi(x/D)$ with $D=\lambda/2$, overlaid with example transfer functions of the ambient medium $H_a(k)$ (low-pass filtering) and of the DFHL $H(k)$ (high-pass filtering).
(b) Comparison between the images of an object with the size $D=\lambda/6$ obtained by \ref{eq:imagedark} (solid line) and \ref{eq:imageblurred} (dashed line) with $n_a =1$ and $k_\text{cutoff}=2k_0$, respectively. The shaded area shows the object $f(x)$ itself. The inset shows the same plot zoomed out in the $x$ axis to show the sinc dependence in $g_a(x)$. (c) Dependence of the subwavelength image visibility $V_\text{sub}$ as defined in \ref{eq:contrast} on the object size given as $k_\text{cutoff}D = 2\pi(k_\text{cutoff}/k_0)(D/\lambda)$.
}
\label{FIG:analytical}
\end{figure}

To see that such high-pass filtering still allows a subwavelength image to be formed, we rewrite \eqref{eq:fourier2D}, reducing it to one dimension (assuming $k_\parallel=k_x$) for simplicity:
\begin{equation}
g(x)=\frac{1}{2 \pi}
\int_{-\infty}^{\infty} dk_x  H(k_x) e^{-i k_x x}
\int_{-\infty}^{\infty} dx' \Pi(x/D) e^{ik_x x'} ,
\label{eq:fourier1D}
\end{equation}
where the object of size $D$ is represented by the rectangular (unit box) function $\Pi(x)$ that assumes unity value for $-1/2<x<1/2$ and zero value elsewhere. In this formalism, the high-pass filtering action of the type-II HMM blocking all the waves below $k_\text{cutoff}$ can be modelled by assuming the transfer function $H(k_\parallel) = 1-\Pi(k_\parallel/k_\text{cutoff})$ (see \ref{FIG:analytical}(a)), which results in the image of the form  
\begin{equation}
g(x)=\Pi\left(\frac{x}{D}\right) - 
\frac{1}{\pi}\left[ \textrm{Si}\left( \frac{D-2x}{4}k_\text{cutoff}\right) + \textrm{Si}\left( \frac{D+2x}{4}k_\text{cutoff}\right) \right],
\label{eq:imagedark}
\end{equation}
where $\textrm{Si}(x)=\int_0^x\textrm{sinc}(t)dt$ is the sine integral function. We see that the high-pass filtering retains the presence of the image in the form of the first term in \eqref{eq:imagedark}, the second term adding some minor background to the image as seen in \ref{FIG:analytical}(b). 
This is unlike the action of the low-pass filtering induced by propagation in some isotropic ambient medium, which can be modelled by similarly assuming the box-type transfer function $H_a(k_\parallel) =\Pi(k_\parallel/n_a k_0)$ and results in 
\begin{equation}
g_a(x)=-
\frac{1}{\pi}\left[ \textrm{Si}\left( \frac{D-2x}{4}n_a k_0\right) + \textrm{Si}\left( \frac{D+2x}{4}n_a k_0\right) \right],
\label{eq:imageblurred}
\end{equation}
which becomes increasingly blurred as the $D n_a k_0$ decreases to values significantly below unity. 

Therefore, we conclude that the image formed by a hyperlens based on the type-II rather than type-I HMM would still be subwavelength as per \eqref{eq:imagedark}, but would have a greatly enhanced contrast compared to the conventional hyperlens because the object-unrelated background, such as signal stemming from the incident light, is blocked. Specifically, the subwavelength image contract, which can be defined as the visibility
\begin{equation}
V_\text{sub}=\lim_{\delta\to0}\frac{|g(D/2-\delta)|-|g(D/2+\delta)|}{|g(D/2-\delta)|+|g(D/2+\delta)|} 
= 1-2{\left[1+\left| \frac{\pi}{\textrm{Si}(k_\text{cutoff}D/2)} -1  \right|  \right]}^{-1},
\label{eq:contrast}\end{equation}
can be seen in \ref{FIG:analytical}(c) to be significant for $k_\text{cutoff}D$ below 2, and to approach unity as the object size decreases. 

We can see that the proposed hyperlens modification is similar in spirit to dark-field microscopy, where background illumination is blocked and only the information coming from the specimen is isolated. Therefore, we call the proposed hyperlens the {\em dark-field hyperlens} (DFHL), and will by contrast refer to the conventional hyperlens as the bright-field hyperlens (BFHL). 

\begin{figure}[tbh!]
\includegraphics[width=0.97\columnwidth]{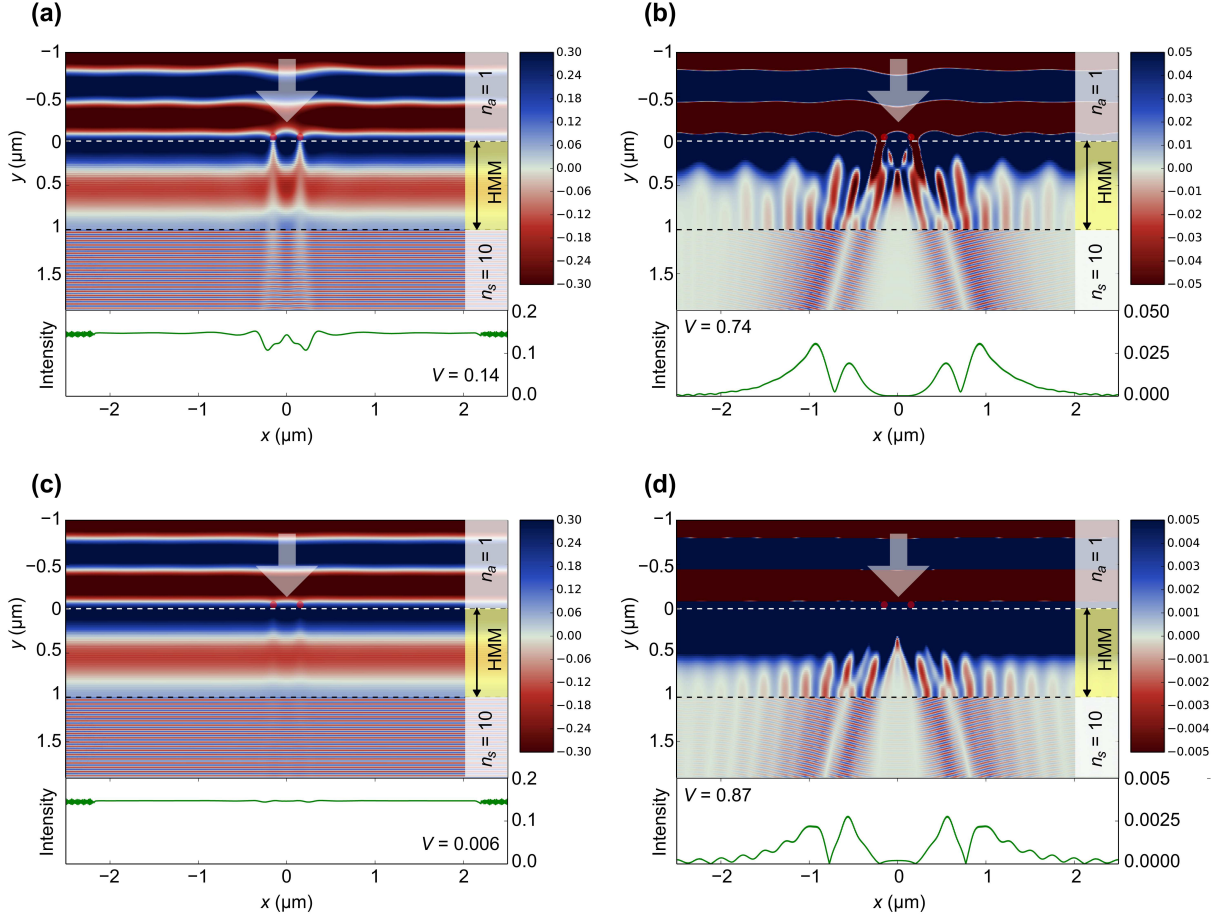}
\caption{
(a--b) Full-wave frequency-domain simulation of a plane wave ($\lambda=715$ nm) impinging on two metallic scatterers (diameter 70 nm, $n = 0.01 + 1.5i$), placed 300 nm apart, in front of (a) conventional hyperlens (BFHL) and (b) proposed hyperlens (DFHL). Both structures are alternating metal-dielectric multilayers containing a total of 100 layers with 10 nm thickness and material parameters $n_m = 0.154 + 1.589i$, $n_d = 1.794$ (type-I HMM) and $n_m = 0.14 + 2.06i$, $n_d = 1.45$ (type-II HMM) for the bright- and dark-field hyperlens, respectively; the corresponding effective-medium dispersion relations are shown in \ref{FIG:Dispersion}. The area behind the lenses contains a high-index medium ($n_s = 10$). (c--d) Same as (a--b) but for dielectric scatterers ($n = 1.5$). The lower plots (green lines) show the $x$-dependence of the field intensity 700 nm behind the lens ($y=1700 \, \text{nm}$).
%
        }
\label{FIG:poormans}
\end{figure}

\section{Planar-geometry ``poor man's'' dark-field hyperlens}

To demonstrate the proposed concept and to compare the functionality of DFHL and BFHL and, we have numerically simulated light propagation in a scenario where two subwavelength metallic scatterers are placed in front of planar hyperlens structures; the scatterers are separated by a distance shorter than the wavelength of the plane wave incident on them. The planar hyperlens geometry is chosen for simplicity, and we call such a structure ``the poor man's DFHL'' because such a design is only able to transmit a subwavelength image of an object some distance without true magnification, by the same token as a slab of metal is dubbed ``the poor man's superlens''\cite{zheludev2008diffraction}. However, despite the obvious drawbacks of this ``toy model'', it is useful as a proof-of-principle that can demonstrate the operating principle of the proposed DFHL without having to regard features brought about by more complicated designs.

%
The results of the simulation are shown in \ref{FIG:poormans}. The structures are multilayers made of alternating metal and dielectric layers, with $2 \times 50$ layers in total. 
All layers have 10 nm thickness, and the BFHL and DFHL structures differ by the material parameters of the metal and dielectric used, with \ref{FIG:Dispersion} showing the corresponding dispersion properties in the effective medium limit (i.e., assuming infinitely thin layers) with and without losses. 

COMSOL finite element software was used for the simulations. We adopt the supercell approach when the simulation domain is periodic in the $x$ direction (tangential to the layers, see \ref{FIG:poormans}) in order to avoid numerical errors arising from metal layers terminating inside perfectly-matched layers (PML) region. (These errors can be significant because the performance of absorbing boundary conditions such as PMLs for the high-$k$ waves existing in HMMs is unstudied and can be very poor.) To prevent artifacts arising from interaction between the scatterers from different supercells, we make the width of the simulation domain 5 {\textmu}m, so any periodicity effects are expected to be negligible. In the $y$ direction (normal to the layers), absorbing boundary conditions are used, with 200 nm thick PMLs placed 1 {\textmu}m away from the multilayer. A normally incident plane wave with wavelength 715 nm is impinging on the hyperlens from the ambient medium with refractive index $n_a = 1$, excited in the simulaiton by surface current boundary condition. On the other side of the multilayer, a substrate with artificially high refractive index ($n_s=10$) is placed so that high-$k$ waves remain propagating, and the subwavelength image pattern can be visualized.

As expected, \ref{FIG:poormans}(a) shows that the BFHL transmits a significant portion of the incident plane wave with $k_\parallel = 0$, which creates a strong background in the image area. The images of the objects are manifest as faint ``shadows'' where the intensity of the background is reduced due to scattering by the objects. We see that even though the two subwavelength images are well-resolved (in agreement with the operating principle of the hyperlens), the image contrast is sufficiently low even for relatively strongly scattering objects such as metallic spheres. To quantify the contrast, we define the bright-field image visibility in the same spirit as 
in \eqref{eq:contrast} as the contrast between the on-image and between-images field intensities: $V_\text{BF} = |I_\text{min} - I_0|/(I_\text{min} + I_0)$, where $I_\text{min}$ is the field intensity at the dip corresponding to each image, and $I_0$ is the intensity at the peak between the dips (at $x=0$), which is almost equal to the background intensity. From \ref{FIG:poormans}(a) it can be recovered that $V_\text{BF}=0.14$.

In contrast, \ref{FIG:poormans}(b) shows that the DFHL reflects the incoming plane wave almost completely, resulting in no background in the image area. Only the subwavelength high-$k$ components of the scattered field are transmitted via coupling to bulk plasmonic waves inside the HMM, predominantly in the form of characteristic cone-like patterns \cite{ishii2013sub} arising because the normals to the isofrequency surface in the dispersion relation \eqref{eq:disp} have a preferred direction [see inset in \ref{FIG:Dispersion}(b)]. Each scatterer produces its own distinct cone pattern, which propagates independently in the hyperlens. As a result, a clearly visible and recognizable subwavelength image pattern is formed at the far end of the poor man's hyperlens, consisting of of two pairs of image points. Comparing the plots of light intensity in the image area ($y = 1700$ nm), we can see that the DFHL produces an image with much lower brightness but with much higher contrast than the BFHL. 
Similarly introducing the visibility as the contrast between the on-image and between-images field intensities, $V_\text{DF} = (I_\text{peak} - I_\text{dip})/(I_\text{peak} + I_\text{dip})$ where $I_\text{peak}$ is the field intensity at the weaker of the two image peaks and $I_\text{dip}$ is the intensity at the dip between the peaks. From \ref{FIG:poormans}(b) we can see that $V_\text{DF}=0.74$, much higher than the corresponding $V_\text{BF}$ obtained with a BFHL.



Let us briefly discuss the key challenges and limitations of the proposed DFHL design. The first challenge consists in the proper choice of the hyperlens thickness in presence of material losses. On the one hand, choosing a thick multilayer would attenuate the useful signal, making the image too weak to be detected. On the other hand, choosing a thin structure would frustrate the filtering properties with respect to the low-$k$ components, so part of the background radiation would still get through and reduce the imaging contrast. This tradeoff is worsened by the fact that the adverse effects of material losses are greater on the high-$k$ waves than on the low-$k$ waves \cite{ourPRB14hmm}. 

Another related tradeoff is the choice of the value of $k_\text{cutoff}$. Choosing a value too close to $n_a k_0$ decreases the attenuation of the low-$k$ components, whereas choosing a higher value prevents part of the high-$k$ information from forming an image and thus decreases the image contrast [see \eqref{eq:contrast} and \ref{FIG:analytical}(b--c)]. So, it follows that lowering losses should be the priority optimization directions for the DFHL, and one may eventually have to resort to active loss compensation using gain media \cite{ni2011loss}.


The third limitation is related to each scatterer producing a cone-shaped pattern within the hyperlens, which poses no difficulty in the analyzed 2D case but requires post-processing to reconstruct the image in the 3D case. For this reason, a hyperlens is usually made to operate in the ``canalization regime'' where the cone is very narrow---ideally, degenerate to be almost line-like, see \ref{FIG:poormans}(a). However, the type-I HMM underlying the BFHL naturally tends to near-canalization regime for sufficiently flat dispersion relation because its topology enforces the group velocity of the lower-$k$ components to point in the direction of $k_\perp$ [see the inset in \ref{FIG:Dispersion} (a)]. In contrast, the type-II HMM in the DFHL does not readily support the canalization regime [see the inset in \ref{FIG:Dispersion}(b)] since the increased curvature of the isofrequency contours near $k_\parallel = k_\text{cutoff}$ causes the group velocity of partial waves in that region to be spread rather than collimated. Furthermore, flattening the dispersion relation (by scaling $\epsilon_\parallel$) also reduces the attenuation of low-$k$ waves, which is determined by $\epsilon_\parallel$, as can be seen by solving \eqref{eq:disp} for $k_\parallel=0$ and yielding $\Im(k_\perp) \approx  k_0\sqrt{-\mathrm{Re}\,\epsilon_\parallel}$ in the limit of small material losses. So, designing the DFHL in the canalization regime remains a challenge, and the proposed example is chosen to operate well outside this regime (see \ref{FIG:poormans}(b)).

However, in spite of these design challenges and tradeoffs, the main advantage of the DFHL, namely its suitability for subwavelength imaging of weak scatterers, is clearly demonstrated in \ref{FIG:poormans}. We note that the metallic particles used in the presented example form rather strong scatterers. Using dielectric particles as objects, as is relevant in label-free biological imaging, would further bring out the advantage of high-contrast imaging facilitated by the DFHL. Indeed, \ref{FIG:poormans}(c--d) shows that the image in the BFHL almost vanishes, becoming indistinguishable against the background of the incident wave, while the DFHL retains the imaging capability. Specifically, the visibilities can be obtained as $V_\text{BF}=0.006$ and $V_\text{DF}=0.87$, i.e., a DFHL produces an image which has more than 140 times better contrast that a BFHL.

%

\section{Cylindrical-geometry dark-field hyperlens}

As discussed in the previous section, the planar-geometry structure only serves as a ``poor man's hyperlens'' in the sense that, while it is capable of transmitting the information encoded in the high-$k$ components of radiation scattered off a subwavelength object through the multilayer thickness, it is not able to perform any actual magnification. The reason is twofold. First, as seen in \ref{FIG:poormans}(b), the image points end up being the same distance apart as the original objects, so such a subwavelength image cannot be further processes by conventional optics. Second, the high-$k$ components traveling in the HMM would still be highly evanescent in any naturally occurring media, which made it necessary to use a fictitious medium with unrealistically high refractive index ($n_s=10$) in order to outcouple these components out of the lens in \ref{FIG:poormans}. 

Both these problems are conventionally solved by employing a cylindrical \cite{jacob2006optical} or spherical \cite{rho2010spherical} geometry. The objects to be imaged are placed at the inner radius of the circular surface, and the high-$k$ waves travel outward towards the outer radius. This gradual geometric transformation, sometimes combined with a gradient imposed on the layers thickness as one moves outward, serve to increase the distance between the image points compared to the distance between the objects, and at the same time, scale the $k$-vector of the waves towards lower values, thus allowing them to be outcoupled and subsequently imaged by conventional optics. 

\begin{figure}
  \includegraphics[width=1.00\columnwidth]{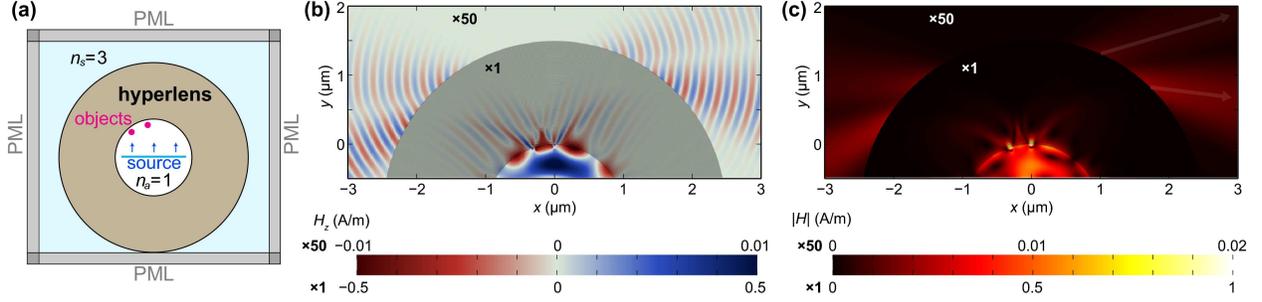}
\caption{The operation of a cylindrical-geometry DFHL: (a) computational schematic and (b--c) numerical results showing (b) field map and (c) field intensity map for a beam incident on two subwavelength scatterers, 70 nm in diameter and 290 nm apart, in front of a DFHL with cylindrical geometry. The structure consists of $2 \times 50$ layers with thickness 15 nm and $n_m = 0.14 + 2.26i$, $n_d = 1.45$; the inner and outer radius of the structure is 1000 and 2500 nm, respectively. To simultaneously show the field before, inside, and after the hyperlens, the color scale of the fields after the hyperlens is magnified by 50.}
  \label{FIG:Hyperlens}
\end{figure}

We have adapted the design principles elaborated in the previous section to the curvilinear geometry, starting with a multilayer structure similar to one used in \ref{FIG:poormans}(b) ($2 \times 50$ layers in total) and making the individual layer thickness 15 nm. The larger layer thicknesses were chosen in order to increase the total thickness of the lens, thereby increasing the magnification factor of the structure. 
Material parameters were also slightly modified in such a way that losses in metal were made slightly larger ($n_m = 0.14 + 2.26i$, $n_d = 1.45$), which was found to increase the quality of the images. The layers form concentric cylindrical shells with inner radius 1000 nm and outer radius 2500 nm. 

The simulation set-up is also quite similar to the previous demonstration of the poor man's hyperlens (see Section 2) except that the ambient space around the lens is now filled with a realistic medium with $n_s=3$ rather than with a fictitious one. Since the structure is now spatially finite rather than infinite, the supercell approach is no longer needed, and hence absorbing boundary conditions are used on all sides. In order to avoid numerical artifacts arising from the interaction of high-$k$ waves with PMLs, complete circular structure is enclosed in the simulation domain [see \ref{FIG:Hyperlens}(a)]; the presence of losses in the HMM structure is expected to prevent the effects of round-trip wave propagation in the structure. 

As before, the objects to be imaged are subwavelength-sized cylinders placed close to the inner surface of the hyperlens. They are illuminated by a beam (FWHM 666 nm) in the $+y$ direction; in the simulation set-up used, this excitation is created by placing a surface current source in middle of the hyperlens, as shown in \ref{FIG:Hyperlens}(a).

\begin{figure}[tbh]
  \includegraphics[width=0.7\columnwidth]{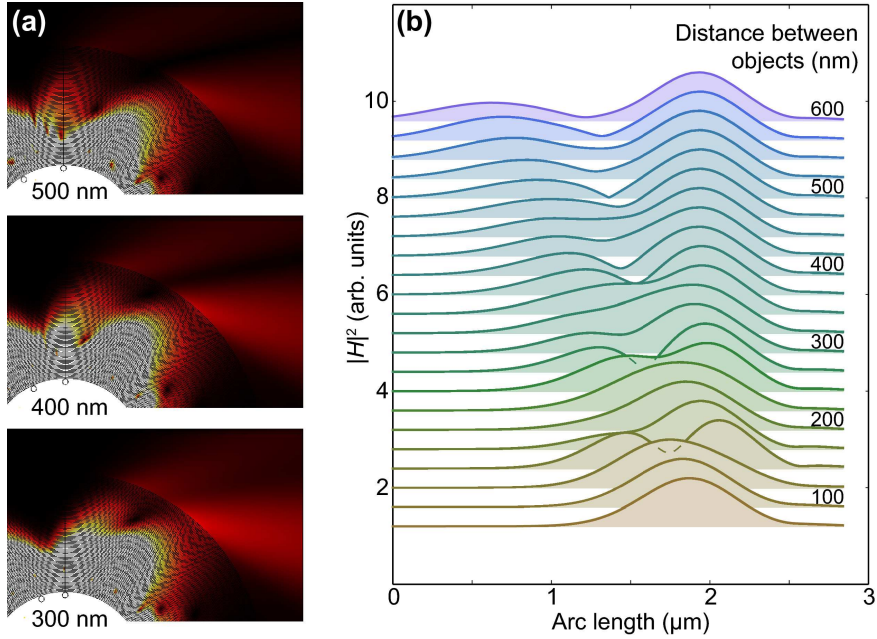}
  \caption{(a) Fragments of intensity maps similar to \ref{FIG:Hyperlens}(b) for three different values (500, 400, and 300 nm) of the distance between the subwavelength scatterers. (b) Dependence of the field intensity at the outer surface of the hyperlens for the distance between objects varying between 75 and 600 nm in 25-nm steps; dots show the theoretical location of the image points in the absence of interference effects between high-$k$ cones.
 }
  \label{FIG:Separation}
\end{figure}

The simulation results are shown in \ref{FIG:Hyperlens}(b--c). The field distribution map (\ref{FIG:Hyperlens}(b)) shows that the operation is quite similar to the planar poor man's hyperlens considered above, with each object giving rise to a cone-like pattern in the hyperlens. However, the high-$k$ waves in these patterns are now transformed as they propagate outward, and can therefore be coupled out of the lens if the output medium has realistic index of refraction.   This can be confirmed in the intensity map (\ref{FIG:Hyperlens}(c)), where it is seen that each object produces a pair of output beams present in the output medium. It can also be seen that the geometric transformation makes the distance between the image beams greater than between objects (e.g. almost 1 micron for 290 nm separation between objects), making the resulting subwavelength image suitable for further processing by conventional optics.

To further characterize the hyperlensing functionality of the proposed structure, we vary the distance between scatterers and analyze how this variation affects the fields in the image area. The results are shown in  \ref{FIG:Separation}. It is seen that bringing the scatterers further apart causes a corresponding increase in the distance between the image points (\ref{FIG:Separation}(a)), with the distance between the centroids of image points nearly proportional to the distance between objects. It can also be seen that  well-resolved images can be formed for objects as close to each other as 175 nm apart (\ref{FIG:Separation}(b)). However, it is seen that the quality of image deteriorates for distances lower than 300 nm, with fluctuating visibility of the image that prevents the two image points from being resolved at some values of inter-object separation. These fluctuations result from the interference of high-$k$ patterns inside the hyperlens. Above 300 nm separation, the subwavelength imaging functionality of the proposed hyperlens is fairly reliable.

The field and intensity profiles seen in \ref{FIG:Hyperlens} and \ref{FIG:Separation} further bring out the drawback of operation outside of canalization regime. The existence of multiple image points per object (or, in the 3D geometry, an image cone) decreases the amount of useful signal and may necessitate the use of image restoration techniques. In addition, the inability to separate the signal from multiple objects inside the hyperlens gives rise to interference effects, which manifest in the image area as distortions. So one of the primary future tasks would be to minimize those interference effects.

Another challenge in the design of the DFHL is that, on the one hand, it should allow propagation of a broad range of high-$k$ waves at its inner surface while having strong attenuation for low-$k$ waves, so that subwavelength image is extracted and separated from background radiation. On the other hand, the waves at the outer surface should have sufficiently low $k$-vectors in order to pass through to the surrounding medium. This tradeoff narrows down the range of high-$k$ waves taking part in the image formation, which limits the imaging resolution according to \ref{FIG:analytical}(b), and gives rise to the requirement that $n_s$ be greater (ideally, much greater) than $n_a$. The account for losses, which affect higher-$k$ waves more strongly, is also detrimental, contributing to narrowing down of the allowed range of high-$k$ waves and lowering the image brightness. Together, these factors place a limit of $2.5\ldots 3$ on the magnification factor of the DFHL achievable by geometrical transformation alone. 

We stress that the presented simulation results only constitute the proof-of-principle for the proposed concept of the DFHL. It is therefore expected that further optimization can improve the DFHL imaging performance. For example, higher magnification and better image quality may be possible with gradient structures where layer thicknesses would vary across the thickness of the lens, although such a structure would be more difficult to manufacture. Another promising optimization direction is to combine two hyperlens structures (BFHL and DFHL) in one device, along the same lines as the combination of a superlens and a hyperlens was used in  \cite{cheng2013breaking}. In such a hybrid device, the DFHL part could perform preliminary magnification along with background radiation filtering, after which the BFHL can perform the main magnification in the canalization regime.




\section{Conclusions and outlook}

In summary, we have proposed a concept for high-contrast subwavelength imaging (hyperlensing) of weakly scattering objects through the use of type-II rather than type-I HMM in the hyperlens, which prevents all propagating waves existing in the object area. The proposed {\em dark-field hyperlens}, so termed because its operating principle is similar to blocking incident light in dark-field optical microscopy\cite{benisty2012dark}, only transmits high-$k$ waves stemming from subwavelength features of the sample (see \ref{FIG:Lenses}). The resulting subwavelength image therefore has a much greater contrast than the one produced by a conventional hyperlens; in the presented numerical example the contrast enhancement by more that two orders of magnitude was demonstrated for subwavelength dielectric scatterers (see \ref{FIG:poormans}). Simulations further confirm the feasibility of the DFHL operation and present proof-of-principle DFHL design with demonstrated subwavelength imaging capability with weakly scattering objects illuminated by external light. This is in contrast with previous studies on the hyperlens \cite{liu2007far,rho2010spherical,cheng2013breaking} where the choice of the sample was intentionally made in such a way as to exclude background radiation.

The proposed concept, {\em dark-field superresolving optical microscopy}, can find many applications in biological imaging because it can be performed without the use of fluorescent markers and/or special sample preparation techniques. As a result, the proposed method can be used to obtain dynamic real-time images of weak-contrast subwavelength objects. The resulting possibility to see and investigate dynamic processes involving subwavelength-sized biological agents and macromolecules (including {\em in vivo} studies) can be enabling to modern life sciences. Finally, the possibility to reverse the operation of a hyperlens, which can be realized by virtue of the time-reversal symmetry of the Maxwell equations, can be used to focus light in a subwavelength scale and selectively excite a subwavelength object. This may bring about an even greater advance in experimental biology, making it possible to actively interfere with nanoscale biological agents while observing them.

\begin{acknowledgement}

This work has received financial support from the People Programme (Marie Curie Actions) of the European Union's 7th Framework (EU FP7) Programme FP7-PEOPLE-2011-IIF under REA grant agreement No.~302009 (Project HyPHONE). One of us (T.R.) thanks the Archimedes Foundation for financial support (Kristjan Jaak scholarship).

\end{acknowledgement}

\begin{suppinfo}

Some movie files can be put here, such as an animation of what happens when the distance between scatterers varies. Our "continuous separation graphs" can be there, too.

\end{suppinfo}


\providecommand*\mcitethebibliography{\thebibliography}
\csname @ifundefined\endcsname{endmcitethebibliography}
  {\let\endmcitethebibliography\endthebibliography}{}

\end{document}